\documentclass[preprint,nofootinbib,tightenlines,showpacs]{revtex4}
\usepackage{amssymb}
\usepackage{amsmath}
\usepackage{lscape}
\usepackage{epsfig}

\def\journal#1, #2, 1#3#4#5, #6{
    {\sl #1~}{\bf #2} (1#3#4#5) #6}

\def\beq{\begin{equation}}
\def\eeq{\end{equation}}
\def\ba{\begin{eqnarray}}
\def\ea{\end{eqnarray}}

\newcommand{\h}{Hamiltonian}

\newcommand{\baa}{\begin{eqnarray}}

\begin{document}

\title{Homolumo Gap from Dynamical Energy Levels }
\author{I. Andri\'c}
\email{iandric@irb.hr}
\author{L. Jonke}
\email{larisa@irb.hr}
\author{D. Jurman}
\email{djurman@irb.hr}
\affiliation {Theoretical Physics Division, Rudjer Bo\v skovi\'c Institute, P.O. Box 180, 10002 Zagreb, Croatia}
\author{H. B. Nielsen}
\email{hbech@nbi.dk}
\affiliation{The Niels Bohr Institute, Copenhagen DK 2100, Denmark}
\begin{abstract}
We introduce a dynamical matrix model where the matrix  is interpreted as a Hamiltonian representing interaction of a bosonic system with a single fermion. We show how a system of second-quantized fermions influences the ground state of the whole system by producing a  gap between the highest eigenvalue of the  occupied  single-fermion states and the lowest eigenvalue of the  unoccupied single-fermion states. We describe  the development of the gap in both, strong and weak coupling regime, while for the intermediate coupling strength  we expect formation of  homolumo "kinks". 
 \end{abstract}

\maketitle
In the complex systems composed of fermions interacting with bosons, 
such as a molecules, nuclei etc.,  appearance of the 
energy gap between highest energy level occupied by fermion and lowest unoccupied
level, so called homolumo gap \cite{jt}, is a well known fact confirmed by experiments as well
as by exact calculations from the first principles in various specific examples.
Some of the  most important properties of the system, stability, 
interaction with another system, size etc., are determined by 
the physics in the neighbourhood of the gap.  
In a more general setting, we are interested in the application of the homolumo effect in the project of Random Dynamics \cite{hf}. There, one starts from the observation that the energies at our disposal are extremely low, compared to the fundamental energy scale, presumably to be identified with the Planck scale. Consequently,  from the fundamental scale point of view, the usual high-energy physics can be described as low-energy excitations in the neighbourhood of the gap. 

The purpose of this letter is to provide a model
which possesses the main features of the interactions involved in the 
production of the homolumo gap and  allows  generalisations
which incorporate other properties such as the appearance of the single level in the gap
or mixing of the densities of the occupied and unoccupied levels.
This letter can be viewed as continuation of our previous work \cite{hl1} which, using different approach,  gives a better insight into mechanism of 
production of the gap. 
In this new approach we would like to confirm that the homolumo gap arises whenever we have a system of fermions and  bosons in interaction, provided that bosons are sufficiently soft to yield to the pressure from fermions, while the details of the model itself seems to be unimportant.

The main assumption of the model  is dynamics of single 
fermion energy levels. This assumption can arise, for example, from the observation  that  the energies of the single electron levels change as the nucleus of a molecule vibrates.
We  show that the \h \  describing the 
system of interacting bosons and fermions can be constructed from very general
assumptions and can be written\footnote{The equations should be understood as matrix equations;  $B$'s are represented by matrices and $f$'s as rows and columns.} as
\begin{equation}\label{h3}
H=H_B+gH_{FB},
\end{equation}
with $H_B$ and $H_{FB}$  given by 
\begin{eqnarray}
H_B=\omega (B^\dagger B+BB^\dagger),\;H_{FB}=f^\dagger B f+f^\dagger B^\dagger f.
\end{eqnarray}
 This form of the \h \  appeared in description of  black holes  \cite{Polch}, mesons and hadrons  in QCD \cite{Aff} and Jahn-Teller effect \cite{jt}.
Within our approach,  the part of the \h \  which describes 
fermion-boson interaction is completely determined, while the 
\h \  describing bosonic degrees of freedom is  a matter of 
choice. 
Our choice is motivated by possible applications in different branches of physics.
As written, the \h \ $H_B$ has been used in description of (a sector within) supersymmetric Yang-Mills theory in four dimensions, two-dimensional quantum gravity and string theory \cite{Alexandrov}, and matrix cosmology \cite{krc}. It is also related to integrable models such as the Calogero model \cite{cal} .
Furthermore, the ground state wave function of $H_B$ corresponds to the 
Gaussian ensemble from the Random Matrix Theory (RMT) \cite{mehta} which has been successfully applied in analyses of the spectra of complex molecules and nuclei, transport properties of disordered mesoscopic systems\cite{guhr}.

We start modelling  the  \h \  by considering  system of the 
interacting fermions described by the  \h \ in which  we include appropriate antisymmetrization of fermionic modes
in analogy with symmetrization of bosonic modes:
\begin{equation}
\label{h1}
H_{FB }={1 \over 2}\sum_{l,k}\left(f_l^{\dagger}f^k-f^k  f_l^{\dagger}\right)X^l_k,
\end{equation}
where matrix $X$ is, for the moment, fixed $N\times N$ hermitean matrix and 
$f^i,f_i^{\dagger}, i=1,\ldots,N$ are fermionic operators satisfying the usual 
anticommutation relations.
Using the appropriate unitary matrix $U$, the matrix $X$ and the fermionic operators 
can be rewritten as  
\begin{equation}\label{diagonalisation}
X^l_k=\sum_{i} U^{\dagger k}_ix_i U^i_l,\;
f^{\dagger}_i=\sum_k d_k^{\dagger}U^{k}_i,\; f^j=\sum_l U^{\dagger j}_l d^l,
\end{equation}
where $x_i$'s are eigenvalues of the matrix $X$. 
Expressed in the new coordinates the \h \ $H_{FB }$ becomes
\begin{equation}\label{h2}
H_{FB }={1 \over 2}\sum_i x_i \left(d^{\dagger}_i d^i-d^i d^{\dagger}_i\right).
\end{equation}
As the  \h \ $H_{FB}$ does not change the number of fermions, the eigenstates  can be chosen with definite number of fermions   
\begin{equation}\label{p1}
\left|\Psi_{i_1,\ldots,i_n}\right.\rangle=d_{i_1}^{\dagger}\cdots d_{i_n}^{\dagger}|0\rangle_f,
\end{equation}
where $|0\rangle_f$ represents fermionic vacuum with property $d_k|0\rangle_f=0$.
As usual, we interpret operator $d_i^{\dagger}$ as creation operator for a fermion at level $i$ 
with associated energy  $x_i$ and $\left|\Psi_{i_1,\ldots,i_n}\right.\rangle$ describes state in which levels 
$i_1,\ldots,i_n$ are occupied while the rest of the levels are unoccupied. 
Therefore, the form of  $H_{FB}$ is completely determined by the requirement that the eigenvalues $x_i$ are interpreted as the energy levels of single fermion.

Next, we  introduce a model where the matrix elements $X_j^i$ are not fixed,
but are random variables, distribution of which is determined by a probability law $P(X_j^i)$. 
Then, we can naturally ask the question about densities of
occupied and unoccupied levels in a certain state of the system.
These are given by the expectation values of the operators
\begin{eqnarray}\label{dens}
\rho^{\mathrm{occ}}(x)\! = \!\!\sum_{i=1}^N\!\delta(x\!-\!x_i)d^{\dagger}_i d^i,
\rho^{\mathrm{unocc}}(x)\! =\!\!\sum_{i=1}^N\!\delta(x\!-\!x_i) d^i d^{\dagger}_i. 
\end{eqnarray}
For the fixed matrix $X$, the densities (\ref{dens}) in a generic eigenstate 
are given by the appropriate sums of the delta functions. 
Particularly, in the ground state of the system, the interval of non-vanishing  density of occupied levels 
lies below the interval of non-vanishing density of  unoccupied levels.
The main consequence of introducing the distribution law  $P(X_j^i)$ is that the sharp delta-functions
profiles are smeared, thus allowing the penetration of the density of the unoccupied levels into the interval 
in which density of occupied levels is nonzero and vice versa. In such situation one expects that the gap between highest occupied level and lowest unoccupied level disappears. We explore this question in the setting where matrix $X$ is dynamical matrix, so that probability law 
$P(X_j^i)$ is a consequence of the quantisation of the dynamical degrees of freedom\footnote{One might say that we approximate fundamentally random physics by more tractable Hamiltonian dynamics. The other point of view would be to attribute the success of RMT to the underlying Hamiltonian dynamics.} of the matrix $X$.
In that case the matrix degrees of freedom contribute to the total energy of the system and to be specific, we assume that self-energy  of the matrix $X$ is determined by the following \h :
\begin{equation}\label{hB}
H_B=\frac{1}{2}\sum_{ij} P^i_j P^j_i+\frac{\omega^2}{2}\sum_{ij} X^i_j X^j_i,
\end{equation}
where $P^i_j=-i{\partial}/{\partial X^j_i}$ is momentum conjugate to $X^j_i$.
After quantisation, $H_B$ can be expressed in terms 
of the appropriate bosonic operators and together with $H_{FB}$, which we now interpret as \h \ of fermion-boson interaction,
constitute the system governed by the \h \ (\ref{h3}). 
The parameter $g$ defines fermion-boson interaction strength and can be taken as positive number without loose of generality, since for negative $g$ we can transform $X\to -X$.

In the following, we are interested in the behaviour of densities (\ref{dens}) in the ground state of the system. In the strong coupling limit  the dominant behaviour of the ground state  is determined by $H_{FB}$.
This means that the suitable fermionic coordinates are the coordinates which diagonalise $H_{FB}$.  
Accordingly, we rewrite the \h \ (\ref{h3}) in terms of  $x_i$'s, 
$U^j_l$'s, $d_i^{\dagger}$'s and $d^j$'s defined by (\ref{diagonalisation}),
expressing derivatives as \cite{Aff}: 
\begin{eqnarray}
{\partial \over \partial X^k_l }=\sum_{i}U^i_k U^{\dagger l}_i{\partial \over \partial x_i}+
\sum_{i,j,m,m\neq i}{U^i_k U^{\dagger l}_m U^m_j \over x_i-x_m}{\partial \over \partial U^i_j}+\sum_{m,n,m\neq n}{U^m_k U^{\dagger l}_n \over x_n-x_m} d_m^{\dagger}d^n.    
\end{eqnarray}
This transformation of coordinates induces a nontrivial measure in the definition of scalar product of the states. 
After performing similarity transformation and defining new \h \ $\tilde{H}=SHS^{-1}$ with $S$ being 
$S=\prod_{i\neq j}\left(x_i-x_j\right)^{1/2}$, the scalar product of new states $|{\tilde\Psi}\rangle=S|\Psi\rangle$ is defined with respect to desirable trivial measure.
Finally, the \h \ (\ref{h3}) can be recast into the following form
\begin{eqnarray}\label{h4}
\tilde{H}=
-{1\over 2}\sum_k {\partial^2 \over \partial x_k^2}
+{\omega^2 \over 2}\sum_k x_k^2+
{g\over 2} \sum_k x_k \left(d^{\dagger}_k d^k-d^k d^{\dagger}_k\right) 
+{1\over 2} \sum_{m,n, m\neq n} {L_m^n L_n^m \over (x_m-x_n)^2},
\end{eqnarray}
where we defined for $n\neq m$
\begin{equation}
L_m^n= U^n_l {\partial \over \partial U^m_l}-\frac{i}{2}\left(d^{\dagger}_m d^n-d^nd^{\dagger}_m\right) .
\end{equation}
The operators $L_m^n=U_m^{k\dagger} J_k^l U_l^n$ are  transformed "angular momentum" generators $J_k^l$ which commute with \h \ (\ref{h3}) and generate unitary group.
Assuming that the last term in the \h \ (\ref{h4}) may be ignored, assumption to be justified latter on, 
the eigenstates of the  \h \ (\ref{h4}) can be written as\footnote{The condition that the eigenstate of the Hamiltonian $H$  is antisymmetric with respect to exchange of two fermions implies that state transformed by aforementioned similarity transformation is  
symmetric with respect to this exchange. By this transformation the matrix degrees of freedom are effectively described as fermions.}:
\begin{equation}\label{st2}
|{\tilde\Psi}\rangle={\tilde\phi}(x_1,...,x_n,y_{1},...,y_{N-n})
 d_1^{\dagger}\cdots d_n^{\dagger}|0\rangle_f ,
\end{equation}
where coordinates  $x_i, i=1,\ldots, n $ correspond to occupied, while coordinates $y_j= x_{n+j}, j=1,\ldots,N-n$ correspond to unoccupied levels.
In the following, we require that function  ${\tilde\phi}$ is antisymmetric 
under the exchange of two indices of the occupied levels and separately 
under the exchange of two indices of the unoccupied levels.
Therefore, the proper state of the system is obtained  by antisymmetrising
the state (\ref{st2}) with respect to exchange of indices of occupied and unoccupied levels.    
The action of the \h \ (\ref{h4}), without the last term, on the state (\ref{st2}) reduces to the action of the \h \ $\tilde{H}_{\mathrm{red.}}$ on the wave function ${\tilde\phi}$, with $\tilde{H}_{\mathrm{red.}}$ given as:
\begin{eqnarray}\label{h5}
\!\tilde{H}_{\mathrm{red.}}\!\!=\!
-{1\over 2}\!\sum_k\! {\partial^2 \over \partial \tilde{x}_k^2}
\!+\!{\omega^2 \over 2}\!\sum_k \!\tilde{x}_k^2
\!-\!{1\over 2}\!\sum_k \!{\partial^2 \over \partial \tilde{y}_k^2}
\!+\!{\omega^2 \over 2}\!\sum_k \!\tilde{y}_k^2,
\end{eqnarray}
where $\tilde{x}_i=x_i+g/2\omega^2,\;\tilde{y}_i=y_i-g/2\omega^2$.
The properly symmetrised ground state of the \h \ $\tilde{H}_{\mathrm{red.}}$ is 
\begin{equation}\label{gsap}
{\tilde\phi}_{\mathrm{gs}}\sim \prod_{i\neq j}\left(\tilde{x}_i-\tilde{x}_j\right)^{1\over 2}
\prod_{i\neq j}\left(\tilde{y}_i-\tilde{y}_j\right)^{1\over 2}
e^{-{\omega \over 2}\sum_i {\tilde{x}}_i^2}
e^{-{\omega \over 2}\sum_j {\tilde{y}}_j^2}. 
\end{equation}
Suppose now that we have exact ground state  $|\Psi_{\rm exact}\rangle$. Due to singularity for $x_i\approx y_j$ of the last term in the \h \ (\ref{h4}), the exact ground state contains the prefactor $\prod_{i,j}\left({{x}_i-{y}_j}\right)$  with the suitable power, as usual in the Calogero-like models \cite{cal}. In the leading order in $g$  the introduction of this prefactor into the state 
(\ref{gsap})  results in multiplication of the state by a constant, and formally the appearance of this prefactor is out of the scope of present approximation. However, the important effect of this prefactor is  
that the contribution to the expectation value of the last term in the \h \ (\ref{h4}) in the exact ground state reduces 
to the principal value integral avoiding singularity.
Using the principal value prescription we can expand the integrand into powers of $g$, showing that this contribution in the state (\ref{gsap}) is of order  $g^{-2}$ relative to the other terms in the \h. 
This justifies our assumption that the last term of the \h \ (\ref{h4}) may be ignored and the bosonic part of the ground state separates into the product of the part depending solely on the occupied levels and the part depending on the unoccupied levels.
Using the usual methods from random matrix theory \cite{mehta}, evaluation of the densities (\ref{dens}) in approximate ground state  gives
\begin{eqnarray}
\label{evalden}
\langle\rho^{occ}(x)\rangle=e^{-\omega \left(x+{g\over 2\omega^2}\right)^2}\sum_{i=1}^{n}
{H^2_i\left(\sqrt{\omega} \left(x+{g\over 2\omega^2}\right)\right)\over 2^i i!\sqrt{\pi}},\\
\langle\rho^{unocc}(x)\rangle=e^{-\omega \left(x-{g\over 2\omega^2}\right)^2}\sum_{i=1}^{N-n}
{H^2_i\left(\sqrt{\omega} \left(x-{g\over 2\omega^2}\right)\right)\over 2^i i!\sqrt{\pi}},\nonumber
\end{eqnarray}
where $H_n(x)$ is Hermite polynomial of order $n$.
\begin{figure}[h]
\centerline{ \epsfig{file=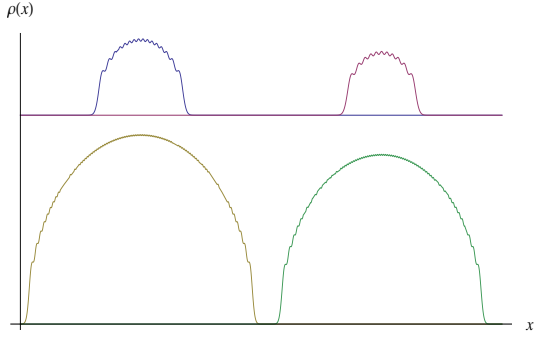,height=4.5cm,width=7cm}}
 \caption{Smoothing out the densities (\ref{evalden});  on the upper graph is plotted $\rho^{occ}(n\!=\!20)$  and $\rho^{unocc}(N\!-\!n\!=\!15)$, and on the lower graph   is plotted $\rho^{occ}(n\!=\!100)$  and $\rho^{unocc}(N\!-\!n\!=\!80)$, with $g/2\omega^2=15$.}
 \label{fig1}
\end{figure}
In the limit $N-n,n \to \infty$, densities (\ref{evalden}) 
reduce to Wigner's semicircle laws, as indicated in Fig.(\ref{fig1}), with centres separated by $g/\omega^2$ in accordance with two-cut solution previously found \cite{hl1}. 
This result shows that in the strong coupling limit we still have well separated densities of occupied and unoccupied levels, although the probability nature of the levels dynamics allows mixing of levels. Furthermore, Eq.(\ref{evalden}) shows that in the case of finite number of levels and finite number of fermions there exist states in the previously found homolumo gap \cite{hl1}.
Note that in precisely this regime, i.e., in the limit of strong interaction, the model defined by (\ref{h3}) was used as a toy model for QCD in the analysis of the spectrum of mesons and baryons \cite{Aff}.

 In the case of weak interaction, up to first order in $g$ we can write the \h \ (\ref{h3})  as
\begin{equation}\label{h3a}
H=e^{iH_{fb}}H_Be^{-iH_{fb}},\end{equation}
where 
\begin{equation}
H_{fb}={g\over 2\omega^2}\sum_{l,k}\left(f_l^{\dagger} f^k-f^k f_l^{\dagger}\right) P_k^l. 
\end{equation}
The eigenstate of the \h \  (\ref{h3a}) is 
\begin{equation}
|\Psi\rangle=e^{iH_{fb}} |\Phi\rangle,
\end{equation}
where $|\Phi\rangle$ is an eigenstate of $H_B$.
The density of occupied eigenvalues in this state is 
\begin{eqnarray}
\langle\rho^{occ}(x)\rangle&=&\langle\Psi|\sum_{i} d_i^{\dagger}d^i\delta(x-x_i) |\Psi\rangle =\nonumber\\
 &=&\sum_{i} \langle\Phi|e^{-iH_{fb}}d_i^{\dagger}d^i\delta(x-x_i)e^{iH_{fb}} |\Phi\rangle.
\end{eqnarray}
Expanding in $g$ up to first order we obtain
\begin{equation}
\langle\rho^{occ}(x)\rangle = \langle\rho^{occ}_0(x)\rangle+{g\over 2\omega^2}\partial_x\langle\rho^{occ}_0(x)\rangle,
\end{equation}
where $\langle\rho^{occ}_0(x)\rangle$ is the density of occupied levels for $g=0$.
Analogously we find 
\begin{equation}
\langle\rho^{unocc}(x)\rangle = \langle\rho^{unocc}_0(x)\rangle-{g\over 2\omega^2}\partial_x\langle\rho^{unocc}_0(x)\rangle.
\end{equation}
The case of weak interaction shows that starting with the system of bosons determined by the 
\h \ $H_B$, introduction of boson-fermion interaction results in the displacement of the 
occupied levels by $-g/2\omega^2$ and unoccupied levels by $g/ 2\omega^2$, whatever these densities are in the case $g=0$. 
\begin{figure}[h]
\centerline{ \epsfig{file=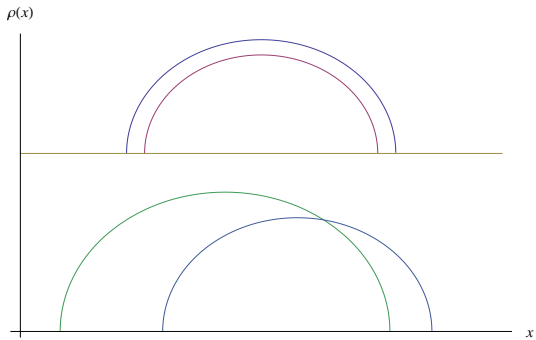,height=4.5cm,width=7cm}}
 \caption{We sketched development of gap in the weak coupling limit for $N-n,n\to\infty$.}
 \label{fig2}
\end{figure}
Moreover, one can show that even to the second order the  boson-fermion interaction results only  in the displacement of the 
of occupied levels by $-g/2\omega^2$ and unoccupied levels by $g/ 2\omega^2$. As this is exactly the result of the strong coupling approximation, one might speculate that this would  be also true for the intermediate $g$.  In that case, in the transition region where densities, although displaced, still overlap, as in  Fig(\ref{fig2}), the total density of levels in the large $N$ limit (the sum of two semicircle distributions) would display two (ignoring the end-points) homolumo 'kinks'.  

The appearance of the gap in our picture depends on  the frequency $\omega$, the interaction strength $g$ and the number of levels $N$ and number of  fermions  $n$. The control over these parameters effectively  controls the size of the gap. A possible disappearance of the gap, observed in  reality, can be  also interpreted within our picture as a consequence of the existence of the several subsystems whose gaps are arranged in appropriate way. Namely, one could introduce different frequencies $\omega_i$ in bosonic part of the model, which would result in different gap for each subsystem, and these could be arranged to overlap.
This could provide a simple model for exploring the insertion of a single level into the gap.

Furthermore,  the  \h \ (\ref{h3}) looks exactly as the \h \ used to derive the linear Jahn-Teller effect \cite{jt}, which says that due to the degeneracy of orbital fermions levels  the symmetry of the molecule is broken in the ground state. Although in description of this effect the  \h \ (\ref{h3}) is defined in configuration rather than in momentum/energy space, the homolumo gap we observe can be interpreted as manifestation of breaking of gauge symmetry, representing  generalisation of Jahn-Teller effect.

In our approach,  the crucial point was use of suitable unitary transformation that enabled us to diagonalize  the model in both strong and weak coupling approximation. On the other hand, using  the conserved  "angular momentum" generators and  appropriate spectrum generating algebra, one should attempt to construct exact eigenstates of the system. We hope to report on the related  results in future publications.

\section*{Acknowledgments}
This work was supported by the Ministry of Science, Education and Sports of the
Republic of Croatia under the contract 098-0982930-2861, and by ESF within the framework of the Research Networking Programme on "Quantum Geometry and Quantum Gravity".


\begin{thebibliography}{}
\bibitem{jt}
H. A. Jahn and E. Teller, Proc. Roy. Soc. London {\bf A161} (1937) 220;
M. Pope and C. E. Swenberg, Electronic Processes in Organic Crystals and Polymers, (2nd ed., Oxford University Press, NY (1999)). 
\bibitem{hf}
   H.~B.~Nielsen,
   ``Dual Strings,''
in ``Scottish Universities Summer School'' (1974) in Sct. Andrews;
D.~L.~Bennett, N.~Brene and H.~B.~Nielsen,
   Phys.\ Scripta {\bf T15} (1987) 158;
C.~D.~Froggatt and H.~B.~Nielsen, \textit{Origin of symmetries}, World Scientific (1991);
C.~D.~Froggatt and H.~B.~Nielsen, Annalen der Physik \textbf{14}, (2005) 115;
see also www.nbi.dk/~kleppe/random/qa/qa.html .
\bibitem{hl1}
I.~Andri\'c, L.~Jonke, D.~Jurman and H.~B.~Nielsen,
  Phys.\ Rev.\  D {\bf 77} (2008) 127701
  [arXiv:0712.3760 [hep-th]].
\bibitem{Polch}
  N.~Iizuka and J.~Polchinski,
  JHEP {\bf 0810} (2008) 028
  [arXiv:0801.3657 [hep-th]].
\bibitem{Aff}
  I.~Affleck,
  Nucl.\ Phys.\  B {\bf 185} (1981) 346.
J.~D.~Lykken,
  Phys.\ Rev.\  D {\bf 25} (1982) 1653.
\bibitem{Alexandrov}
  S.~Alexandrov,
  arXiv:hep-th/0311273.
\bibitem{krc}
J.~L.~Karczmarek and A.~Strominger,
  JHEP {\bf 0404} (2004) 055
  [arXiv:hep-th/0309138].
\bibitem{cal}
F. Calogero, 
J. Math. Phys. {\bf 10} (1969) 2197; M. A. Olshanetsky, A. M. Perelomov,  
Phys. Rept. {\bf 94} (1983) 313;
I.~Andri\'c and D.~Jurman,
JHEP {\bf 0501} (2005) 039
[arXiv:hep-th/ 0411034].
\bibitem{mehta}
M. L. Mehta, "Random Matrices", (2nd Edition, Academic Press, NY(1991)).
\bibitem{guhr}
T. Guhr, A. Mueller-Groeling, H. A. Weidenmueller, 	Phys. Rept. {\bf 299} (1998) 189.



\end{thebibliography}
\end{document}